%% file: 0-main.tex
\documentclass[10pt,conference]{IEEEtran}
\IEEEoverridecommandlockouts

\usepackage{threeparttable}
\usepackage{listings}
\usepackage{tikz}

\usepackage{amsmath,amssymb,amsfonts}
\usepackage{algorithmic}
\usepackage{graphicx}
\usepackage{textcomp}
\usepackage{subcaption}
\usepackage{xcolor}
\usepackage{colortbl}
\usepackage[switch]{lineno}
\usepackage{color}
\usepackage{booktabs}
\usepackage[T1]{fontenc}
\usepackage{enumitem}
\usepackage[many]{tcolorbox}
\usepackage{url}
\usepackage{cite}

\usepackage{listings}
\usepackage{tcolorbox}
\usepackage{multirow}
\usepackage{balance}
\usepackage{hyperref}

\UseRawInputEncoding

\definecolor{VeryLightGray}{gray}{0.8}
\newcommand{\red}[1]{{\textcolor{red}{#1}}}
\newcommand{\Space}[1]{}

\definecolor{lightred}{rgb}{1, 0.90, 0.90}

\lstdefinelanguage{myJava}[]{Java}{
  morekeywords={},
  morecomment=[f][\color{blue}]{@@},     
  morecomment=[f][\color{red}]-,         
  morecomment=[f][\color{ForestGreen}]+, 
  morecomment=[f][\color{magenta}]{---}, 
  morecomment=[f][\color{magenta}]{+++},
  morecomment=[f][\color{magenta}]{//},
  morecomment=[f][\color{magenta}]{/*},
  morecomment=[f][\color{magenta}]{\ \ \ \ //},
  morecomment=[f][\color{magenta}]{\ \ //},
  keywordstyle=\color{blue},
  commentstyle=\color{gray},
  stringstyle=\color{purple},
  numbers=left,
  basicstyle=\scriptsize\ttfamily,
  numbersep=6pt,
  numberstyle=\tiny\ttfamily,
  breaklines=true,
  escapeinside={(*@}{@*)},
  showstringspaces=false,
  xleftmargin=8pt,
}

\lstdefinelanguage{myC}[]{C}{
  morekeywords={},
  morecomment=[f][\color{blue}]{@@},     
  morecomment=[f][\color{red}]-,         
  morecomment=[f][\color{ForestGreen}]+, 
  morecomment=[f][\color{magenta}]{---}, 
  morecomment=[f][\color{magenta}]{+++},
  morecomment=[f][\color{magenta}]{//},
  morecomment=[f][\color{magenta}]{/*},
  morecomment=[f][\color{magenta}]{\ \ \ \ //},
  morecomment=[f][\color{magenta}]{\ \ //},
  keywordstyle=\color{blue},
  commentstyle=\color{gray},
  stringstyle=\color{purple},
  numbers=left,
  basicstyle=\scriptsize\ttfamily,
  numbersep=6pt,
  numberstyle=\tiny\ttfamily,
  breaklines=true,
  escapeinside={(*@}{@*)},
  showstringspaces=false,
  xleftmargin=8pt,
}

\AtBeginDocument{%
  \providecommand\BibTeX{{%
    \normalfont B\kern-0.5em{\scshape i\kern-0.25em b}\kern-0.8em\TeX}}}

\def\BibTeX{{\rm B\kern-.05em{\sc i\kern-.025em b}\kern-.08em
    T\kern-.1667em\lower.7ex\hbox{E}\kern-.125emX}}

\begin{document}
\begin{sloppypar}
\title{Who is in Charge here? Understanding How Runtime Configuration Affects Software along with Variables\&Constants}

\author{
\IEEEauthorblockN{Chaopeng Luo\IEEEauthorrefmark{1}, Yuanliang Zhang\IEEEauthorrefmark{1}\IEEEauthorrefmark{3}\thanks{\IEEEauthorrefmark{3} Chaopeng Luo and Yuanliang Zhang are co-first authors.} \thanks{\IEEEauthorrefmark{4} Haochen He and Shanshan Li are the corresponding authors.}, Haochen He\IEEEauthorrefmark{2}\IEEEauthorrefmark{4},Zhouyang Jia\IEEEauthorrefmark{1}, \\Teng Wang\IEEEauthorrefmark{5}, Shulin Zhou\IEEEauthorrefmark{1}, Si Zheng\IEEEauthorrefmark{1}, Shanshan Li\IEEEauthorrefmark{1}\IEEEauthorrefmark{4}}
\IEEEauthorblockA{\IEEEauthorrefmark{1}National University of Defense Technology, Changsha, China}
\IEEEauthorblockA{\IEEEauthorrefmark{2}Key Laboratory of Satellite Information Intelligent Processing and Application Research, Beijing, China}
\IEEEauthorblockA{\IEEEauthorrefmark{5}National Innovation Institute of Defense Technology, Beijing, China}
Email:{\{luochaopeng18, zhangyuanliang13, hehaochen13, jiazhouyang, wangteng13, zhoushulin\}@nudt.edu.cn}\\
{si.zheng1009@gmail.com},{shanshanli@nudt.edu.cn}
}

\maketitle

\begin{abstract}
    Runtime misconfiguration can lead to software performance degradation and even cause failure. It is usually caused by invalid parameter values set by users. Developers typically perform sanity checks during the configuration parsing stage to prevent invalid parameter values. However, we discovered that even valid values that pass these checks can also lead to unexpected severe consequences. Our study reveals the underlying reason: the value of runtime configuration parameters may interact with other constants and variables when propagated and used, altering its original effect on software behavior. Consequently, parameter values may no longer be valid when encountering complex runtime environments and workloads. Therefore, it is extremely challenging for users to properly configure the software before it starts running.

    This paper presents the first comprehensive and in-depth study (to the best of our knowledge) on how configuration affects software at runtime through the interaction with constants, and variables (\textbf{PCV Interaction}). Parameter values represent user intentions, constants embody developer knowledge, and variables are typically defined by the runtime environment and workload. This interaction essentially illustrates how different roles jointly determine software behavior. In this regard, we studied 705 configuration parameters from 10 large-scale software systems. We reveal that a large portion of configuration parameters interact with constants/variables after parsing. We analyzed the interaction patterns and their effects on software runtime behavior. Furthermore, we highlighted the risks of PCV interaction and identified potential issues behind specific interaction patterns. Our findings expose the "double edge" of PCV interaction, providing new insights and motivating the development of new automated techniques to help users configure software appropriately and assist developers in designing better configurations.
\end{abstract}


\input{1-intro}

\input{2-methodology}
\input{3-pattern}
\input{4-effect}

\input{5-problem}
\input{7-threats}
\input{8-related}
\input{9-conclusion}
\input{10-acknowledege}

\end{sloppypar}
\bibliographystyle{IEEEtran}
\bibliography{10-ref}
\end{document}

%% file: 1-intro.tex
\section{Introduction}

\begin{figure}
    \centerline{\includegraphics[width=0.48\textwidth]{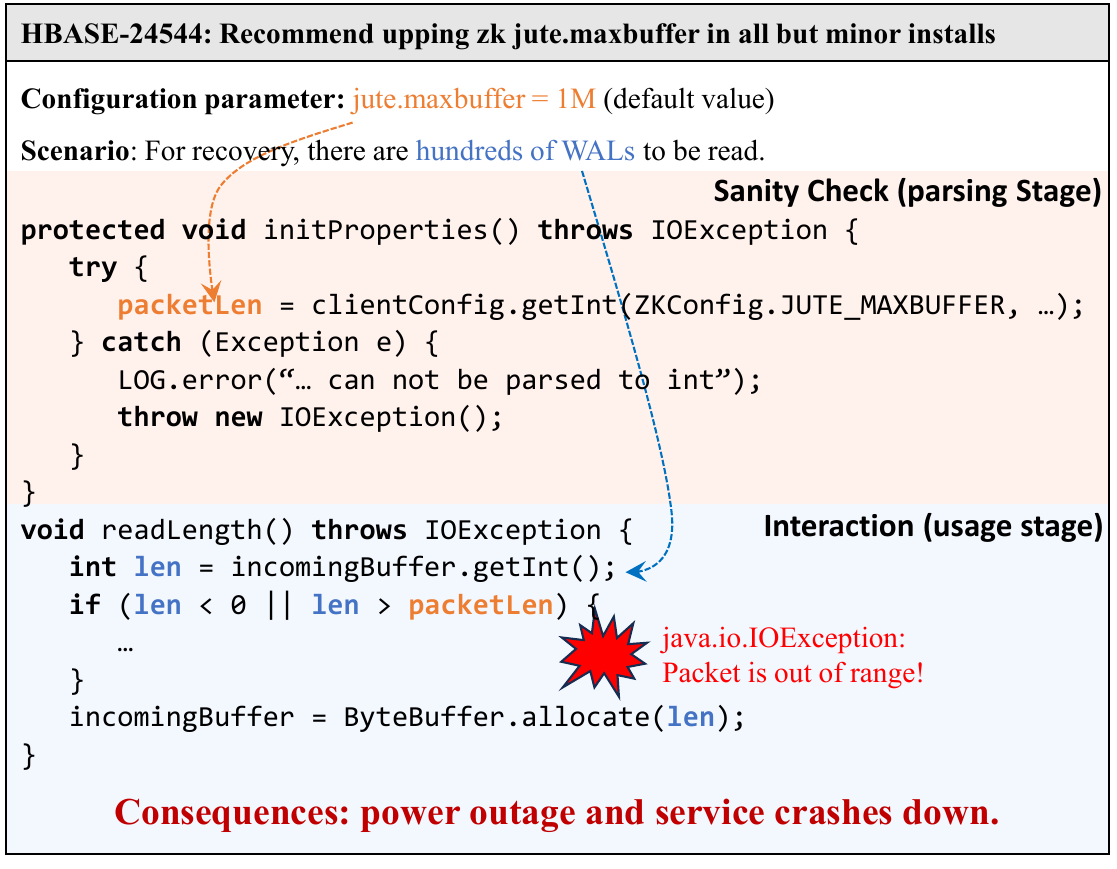}}
    \vspace{-5pt}
    \caption{An example of a valid parameter value causes software crash during interaction.}
    \label{fig:intro}
    \vspace{-15pt}
\end{figure}

Runtime configuration is used at deployment time to port a software system to accommodate different environments and workloads without re-compiling the software~\cite{zhang2021evolutionary, wang2023understanding}.
Despite the convenience, configuration issues have also become one of the major problems of large-scale software systems~\cite{yin2011empirical, xusosp13, xu2015hey}.
For instance, in October 2021, An internal configuration error caused Facebook, Instagram, and WhatsApp to go offline across the globe, affecting nearly 200 million daily users \cite{2021fbWhatsApp}. 
\Space{According to Google's 2023 Cloud Security Threat Report, configuration errors have emerged as the second-largest factor contributing to security incidents in Google Cloud, accounting for a significant 19\% \cite{2023google}.
Additionally, in popular Facebook products, failures related to configuration account for a significant 16\% of service disruptions \cite{tang2015holistic}.}

Many works \cite{xu2016early, yin2011empirical, xusosp13, li2021challenges} find that misconfiguration can be the culprit, and developers constantly add sanity checkers to check basic syntax constraints (e.g., not-NULL, value type) \cite{zhang2021evolutionary}.
However, these checks usually only work during configuration parameter parsing, which is the very early stage of the configuration usage lifecycle. \textbf{After parsing, the configuration parameter will interact with the program in various ways to achieve different goals (e.g., allocating resources and adapting the environment). }

While the complex interaction is a black box to the users, making users hard to accurately predict the effect of configuration, thereby introducing configuration-related issues.  
In fact, many cases \cite{HBASE-21000, HBASE24544, hbase12971, hbase19660, HDFS-12603, mysqlstack} have shown that syntactically or even semantically valid parameter values can lead to serious consequences.

Figure \ref{fig:intro} shows an example \cite{HBASE24544} from HBASE that a valid parameter value may cause a software crash due to interaction with \Space{variable (recording runtime environment/workload)} program workload. The parameter \emph{jute.maxbuffer} restricts the size of data in a single \emph{znode}. In this case, a user sets it with the default value of 1MB (which is the best out-of-box value). However, when a \emph{regionServer} manages hundreds or even thousands of regions, the data volume from backed-up Write-Ahead Logs (WALs) during recovery can reach hundreds or even thousands. The 1MB buffer cannot handle such a huge workload, causing cluster crashes. After inspecting the source code, this parameter value interacts with the variable "\emph{len}" (representing the workload). If the variable's value exceeds the configured threshold, a runtime exception will be thrown. The reporter suggested resolving the problem by increasing the default value of this parameter. However, such an approach cannot solve the problem completely as the software will face an even larger workload.

Considerable effort has been made to address configuration issues \cite{huang2015confvalley, zhang2014encore, xusosp13, wang2023understanding, he2020cp-detector, sun2020testing}. 
However, Existing work mainly focuses on detecting invalid configuration values or flawed configuration parsing code. 
They lack a comprehensive and deep understanding of how configuration affects software behavior, especially in terms of the interaction between configuration parameters and other program constants or variables. Configuration parameter values represent user intention, while constants incorporate developer knowledge and variable values are usually determined by runtime environment/workload. 
Understanding such interactions can help developers learn good practices and realize potential risks that configuration may cause.
Additionally, it can help users set parameter values with greater precision (beyond simply "valid") by elucidating the impact of these parameters on performance and reliability.

In this paper, we conduct the first comprehensive analysis (to the best of our knowledge) of the interaction of parameter \& constant \& variable (\textbf{PCV Interaction)} at the source code level to understand how configuration affects software at runtime. We organize the study by answering four research questions: 

\begin{itemize}[leftmargin = 2em]
\item\textbf{RQ1:} Are PCV interactions common in software? 
\item\textbf{RQ2:} What are the types and patterns of PCV interactions? 
\item\textbf{RQ3:} How do PCV interactions take effects on software? 
\item\textbf{RQ4:} What are the potential problems behind PCV interactions?
\end{itemize}

To achieve this, we study 705 configuration parameters from 10 extensively used, large-scale software systems, tracing parameter parsing, propagation, and usage in the source code and manually analyzing the results. We find that a large portion of configuration parameters (66.4\%) will interact with other constants/variables during propagation and usage, and the interactions can have a great impact on software performance and reliability. 
We conclude seven types of interaction patterns (\S \ref{section: PCV Pattern}).
These interactions can affect software behaviors in four main aspects: misconfiguration prevention, environment adaption, workload adaption, and fault tolerance (\S \ref{section: Interaction Effect}).
We further analyze and identify potential problems behind four main interaction patterns, providing future research directions with actionable suggestions for each identified issue (\S \ref{section: potential problems}). 


Overall, this paper makes the following contributions:
\begin{itemize}[leftmargin = 2em]
\item\textbf{New perspective of how configuration affects software at runtime.} We study how configuration affects software at runtime from the view of PCV interaction, which introduces three different roles: user intention, developer knowledge, and runtime environment/workloads. We comprehensively analyze the patterns, effects, and potential problems of PCV interaction.

\item\textbf{Findings and insights.} We present novel findings and insights derived from our study. \Space{The findings are summarized in Table \ref{tab: summary of findings}.} Our findings and insights reveal the "double-edge" of PCV interactions, which can help developers and users to better design and use configuration. Also, we point out several research directions for future configuration automated tools.

\item\textbf{Dataset.} We release our dataset in this paper for future research in this area. Our dataset can be used for extensive study and tool assessment. The dataset is available at: https://github.com/PCVAnonymous/PCVStudy
\end{itemize}

%% file: 2-methodology.tex
\section{study methodology}\label{section: methodology}

To study the interaction between constants and variables in configurations and programs. We aim at source code level study, which is the most fine-grained way to reveal the interaction relationship. We start with configuration parameters and then find their related code snippet and background knowledge (commits and issues). 


\subsection{Selecting Software and Configuration Parameters}\label{section: config selection}

    \paragraph{\textbf{Target software}}
    To ensure the representation of our study results, we choose 10 software systems from different domains, including database, file system, data processing, and web server, as shown in Table \ref{tab: target software}.
    We selected these software systems because they 1) are mature, widely-used software systems in different fields, 2) have a large number of configuration parameters with detailed user manuals and documentation, and 3) are actively developed with well-organized GitHub repositories and bug databases.
    
    \paragraph{\textbf{Configuration parameters}}
    
    Manually studying and analyzing all parameters is impractical. We first collect parameters in the default configuration file as they are visible to all users (some parameters are hidden in source code and only used by experts or developers) and get 3523 parameters. Subsequently, we randomly sample 20\% of them and finally study 705 parameters across all software systems. The distribution is shown in Table~\ref{tab: target software}.

    \Space{
    Our preliminary investigation revealed certain types of configuration parameters usually have few interaction with program variables and constants and will be directly used after parsing (e.g., path). We excluded parameters with specific keywords (\verb|key|, \verb|dir|, \verb|url|, \verb|port|, \verb|name|, \verb|address|, \verb|path|, \verb|user|, \verb|socket|, \verb|password|, \verb|host|, \verb|id| and \verb|file|) in their name from consideration. We filter out \red{1331} parameters and leave \red{3523} parameters (\#P in Table~\red{tab: target software}). Subsequently, we randomly sampled 20\% of these parameters for each software and finally studied 705 parameters
    (\#Ps in Table~\ref{tab: target software}) across all software systems.}

\begin{table}
    \caption{Software selected in our study}
    \label{tab: target software}
    \vspace{-5pt}
    \begin{center}
    \begin{tabular}{lllcc}
    \toprule
    \textsc{\textbf{Software}} & \textsc{\textbf{Desc.}} & \textsc{\textbf{Lang.}} 
    & \textsc{\textbf{\#P}} & \textsc{\textbf{\#Ps}}\\
    \midrule
    Httpd &Web server & C  & 557 & 111 \\
    PostgreSQL & Database & C  & 251  &50 \\
    Nginx & Proxy server & C  & 480 &96 \\
    MySQL  & Database & C++  & 390 &78 \\
    HBase  & Database &Java  & 174 &35 \\
    Hive &  Database &Java & 484 &97\\
    HDFS & File system &Java & 463 &93 \\
    Yarn & Resource manager &Java & 450 &90 \\
    MapReduce & Data processing &Java & 168 &34 \\
    ZooKeeper & Config manager &Java & 154 &21\\
    \midrule
    Total &- & - &3523 &705\\
    \bottomrule
    \end{tabular}
\end{center}
\vspace{-20pt}
\end{table}

\subsection{Taint Analysis of Configuration Parameters}\label{section: taint analysis}

In order to find the PCV code snippets, we need to trace the propagation and utilization of configuration parameters within the source code. We leverage two open-source taint analysis tools for: Conftainter\cite{conftainter} for C and C++ software and CFlow\cite{cflow} for Java software. 

Typically, the lifecycles of parameters in source code include three main stages, i.e., parsing, propagation, and usage. After setting parameters in command lines or parameter files, the values will be parsed and stored in config variables. If users do not set the value, the default value will be used in the program. The parameter value will be propagated along data-flow using the Use-Define Chain, arithmetic operations (intra-procedural), or referenced by a function (inter-procedural). Previous work predominantly concentrated on the propagation of configurations within the software itself (including Conftainter and Cflow). After the parameter propagates through a series of stages, it will ultimately be utilized in a conditional branch or be referenced by an external function (e.g., syscall and STL function), and then we will stop tracing. 


\subsection{Analyzing Interaction Cases. \textbf{(RQ1)}} \label{section: data collection}

\Space{
 \begin{center}
     \noindent\fbox{
         \parbox{.97\linewidth}
         {
             \textbf{RQ1:} Are PCV interactions common in software?
         }
     }
 \end{center}
 }

The interaction of variables in a program is natural. \textbf{The criterion of PCV interaction is whether the configuration value will be affected by other program constants or variables before it is finally used and takes effect.} In this paper, we determine whether a PCV interaction exists by checking whether the configuration parameter engages in arithmetic or logical operations with constants or variables during its propagation and usage. 

We collect the statement where this interaction occurs. 
If a configuration parameter undergoes multiple interactions during propagation, we will collect all statements where such interactions occur. For a configuration parameter that doesn't undergo any interactions, we consider it as direct usage. Finally, we collected 851 cases (including both interactions and direct usages) for 705 parameters. We manually study each case. We use \texttt{git-blame} to find out the commit that introduces the code and study corresponding issues on JIRA \cite{jira} or GitHub \cite{github}. We analyze the issue description, and developers' discussion along with source code to understand the background and knowledge behind the interaction.


We find only about one-third (237 / 705 = 33.6\%) of parameters can independently affect software behavior without conducting PCV interactions. 


\vspace{-20pt}
\begin{center}
    \noindent\begin{tcolorbox}
        \parbox{.97\linewidth}
        {
            \textbf{Finding 1: Only one-third (33.6\%, 237 / 705) of parameters can independently affect software behavior by propagating and using its value faithfully.} The rest of the parameters will interact with constants or variables.
        }
    \end{tcolorbox}
\end{center}

%% file: 3-pattern.tex
\section{PCV Interaction Pattern \textbf{(RQ2)}}
\label{section: PCV Pattern}

\Space{
 \begin{center}
     \noindent\fbox{
         \parbox{.97\linewidth}
         {
             \textbf{RQ2:} What are the types and patterns of PCV interactions?
         }
     }
 \end{center}
 }

To better understand the interaction relationships of PCV, we analyze the code snippets of PCV interactions. There are four main interaction types: 1) Parameters work independently ($P$), 2) Parameters interact with Constants ($P\&C$), 3) Parameters interact with Variables ($P\&V$) and 4) Parameters interact with Constants and Variables ($P\&C\&V$). We further subdivide them and get seven interaction patterns as shown in Table~\ref{tab: pattern of interaction}. We collect 851 data points from the sample of 705 parameters, the statistical results are shown in Table~\ref{tab: pattern distribution}.

\begin{table}[htbp]
    \caption{The patterns of PCV Interaction}
    \label{tab: pattern of interaction}
    \small
    \begin{center}
    \begin{threeparttable}
    \resizebox{\linewidth}{!}{
    \begin{tabular}{cllc}
    \toprule
    \textsc{\textbf{Type}}&\textsc{\textbf{Pattern}}&\textsc{\textbf{Dominant Relation}} & \textsc{\textbf{Abbr.}}\\
    \midrule
   \multirow{1}{*}{P} & $\circ$ = $p$        & P is self-dominant          & $P$\\
   \cmidrule{1-4}
   \multirow{2}{*}{P\&C}& $P=f(p,c)?p:\phi$     & P is constrained by C  & $\textit{\text{C}}\rightarrow\textit{\text{P}}$\\
   & $\circ$ = $g(p,c)$   & P, C take equivalent place     & $\textit{\text{P -- C}}$\\
   \cmidrule{1-4}
   \multirow{3}{*}{P\&V}& $P=f(p,v)?p:\phi$     & P is constrained by V  & $V\rightarrow P$\\
   & $V=f(p,v)?v:\phi$     & P restrains V       & $P \rightarrow V$\\
   & $\circ$ = $g(p,v)$   & P, V take equivalent place       & $\textit{\text{P -- V}}$ \\
   \cmidrule{1-4}
   \multirow{2}{*}{P\&C\&V}& $\circ$ = $g(p,c,v)$ & P, V, C take equivalent place  & $\textit{\text{P--V--C}}$\\
   & $P\&C \wedge P\&V$ & Mixed  & / \\
    \bottomrule
    \end{tabular}
    }
    \begin{tablenotes}
        \scriptsize
        \item[1] f() is a comparison function and g() is an arithmetic function.  
        \item[2] $\phi \in \{p, c, v, g(p,c), g(p,v), \textit{\text{ThrowException}} \}$
   \end{tablenotes}
   \end{threeparttable} 
    \end{center}
    \vspace{-10pt}
\end{table}

\subsection{Parameter works independently (P)}
\label{section: P}

If a configuration parameter, during the process from being parsed to the usage phase, does not interact with other variables or constants, we consider the configuration parameter is independently affecting software behavior. Note that if multiple parameters interact with each other (also known as parameter dependency \cite{cdep}), the interacted parameters belong to this type because they all represent user-set configuration. Figures \ref{fig: hbase p} illustrate an example of configuration parameters independently affecting software behavior. Before being called by external functions (\emph{defaultPool()}), \emph{opThreads} do not interact with any variables or constants.

\subsection{Parameter interacts with Constant (P\&C)}
\label{section: PC}

\paragraph{\textbf{Parameter is constrained by Constant (C$\rightarrow$P)}} 
In this scenario, constants serve to constrain the range of parameters. If the user-defined parameter values fall outside this range, the program will handle this situation. 

\begin{figure*}[htbp]
	\centering
    \begin{subfigure}{0.32\linewidth}
		\centering
		\includegraphics[width=\linewidth]{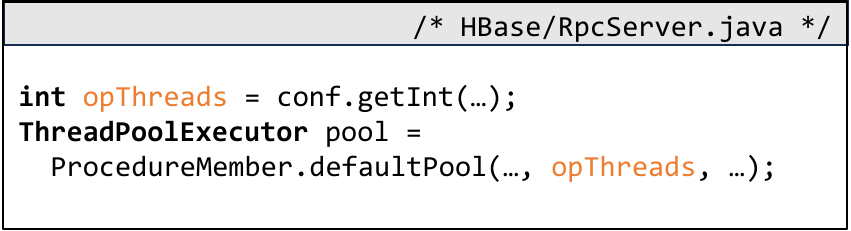}
		\caption{P}
		\label{fig: hbase p}
    \end{subfigure}
    \begin{subfigure}{0.32\linewidth}
		\centering
		\includegraphics[width=\linewidth]{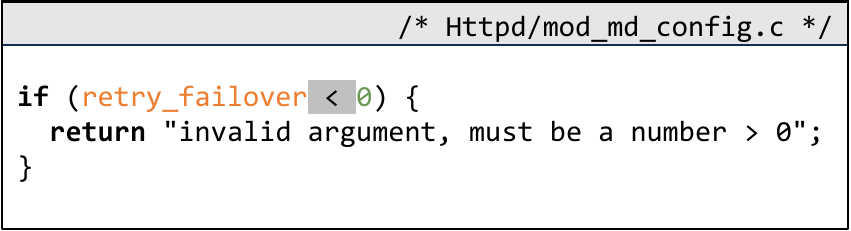}
		\caption{C$\rightarrow$P}
		\label{fig: httpd c->p}
    \end{subfigure}
    \begin{subfigure}{0.32\linewidth}
		\centering
		\includegraphics[width=\linewidth]{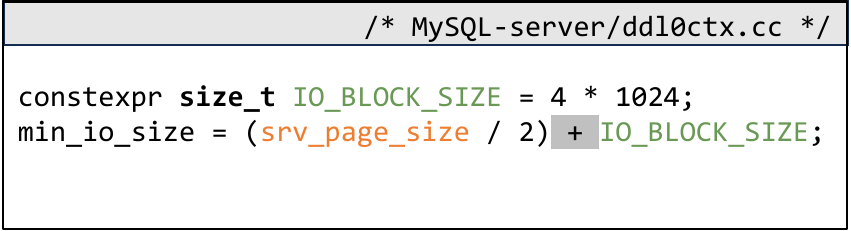}
		\caption{P-C}
		\label{fig: mysql p-c}
    \end{subfigure}
    
    \begin{subfigure}{0.33\linewidth}
		\centering
		\includegraphics[width=\linewidth]{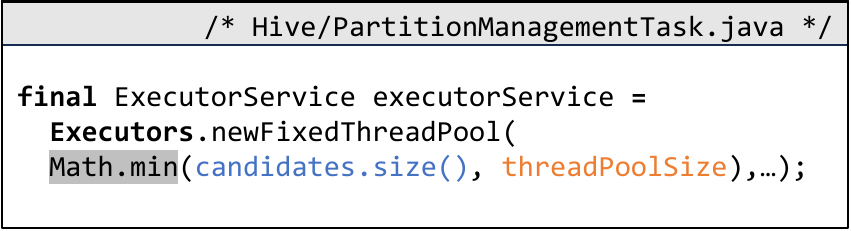}
		\caption{V$\rightarrow$P}
		\label{fig: hive v->p.pdf}
    \end{subfigure}
    \begin{subfigure}{0.33\linewidth}
		\centering
		\includegraphics[width=\linewidth]{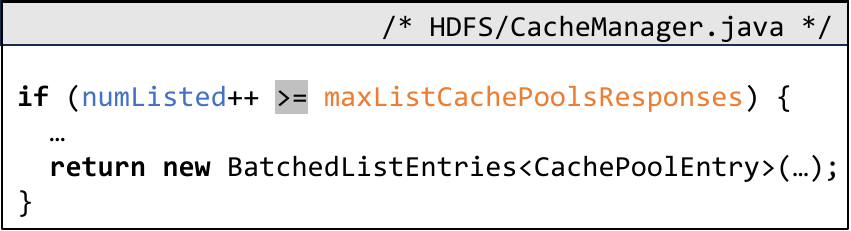}
		\caption{P$\rightarrow$V}
		\label{fig: hdfs p->v}
    \end{subfigure}
    
    \begin{subfigure}{0.33\linewidth}
		\centering
		\includegraphics[width=\linewidth]{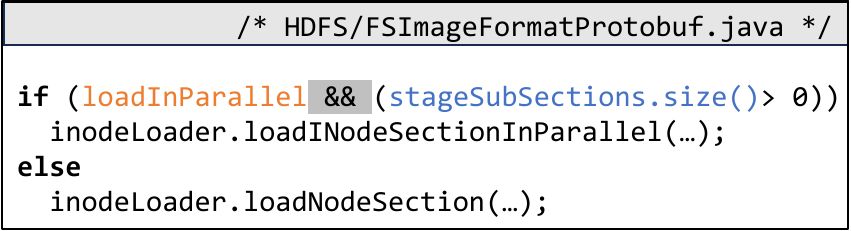}
		\caption{P-V}
		\label{fig: hdfs p-v}
    \end{subfigure}
    \begin{subfigure}{0.33\linewidth}
		\centering
		\includegraphics[width=\linewidth]{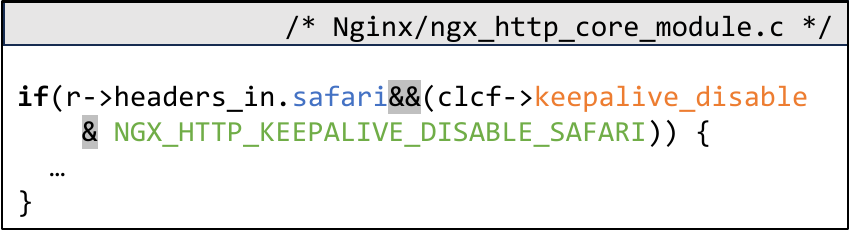}
		\caption{P-C-V}
		\label{fig: nginx pcv}
    \end{subfigure}
    
    \caption{Examples of PCV interaction patterns. Parameters, variables, and constants are marked in orange, green, and blue respectively, and the interactions code are highlighted in gray.}
    \label{figure: PCV interaction}
    \vspace{-15pt}
\end{figure*}

\paragraph{\textbf{Parameter and Constant take equal place (P$-$C)}} 
In this scenario, parameters and constants collaboratively determine (denoted as "$-$") the value of another variable or the execution path of the branch through numerical computations, devoid of any constraints between them. Taking Figure \ref{fig: mysql p-c} as an example, 
In other words, the parameter (\emph{srv\_page\_size}) and the constant (\emph{IO\_SIZE}) jointly determine the size of I/O space (\emph{min\_io\_size}) with equivalent status.

\subsection{Parameters interacts with Variables (P\&V)}
\label{section: PV}

\paragraph{\textbf{Parameter is constrained by Variable(V$\rightarrow$P)}}
In this scenario, the parameter value will be compared with the variable value. Using Figure \ref{fig: hive v->p.pdf} as an illustration. If the parameter \emph{threadPoolSize} exceeds the value of \emph{candidates.size()}, the variable will overwrite the parameter value and determine the size of the thread pool.

\paragraph{\textbf{Parameter restrains Variable (P$\rightarrow$V)}}
Parameters serve as thresholds at conditional branches, constraining the values of variables. 
Figure \ref{fig: hdfs p->v} illustrates an example from HDFS. The parameter \emph{maxListCachePoolsResponses} controls the number of cache pools that the \texttt{NameNode} will send over the wire in response to a \texttt{RPC}, i.e., \emph{numListed}. 

\paragraph{\textbf{Parameter and Variable take equal place  (P$-$V)}} 
Variables and parameters can also engage in interactions similar to \emph{P$-$C}, where they take equal place. 
In Figure \ref{fig: hdfs p-v}, the program runs the parallel function in L4 only if the parameter and variable values are true.

\begin{table}
    \caption{Data distribution of seven interaction patterns}
    \label{tab: pattern distribution}
    \footnotesize
    \begin{center}
    \resizebox{\linewidth}{!}{
    \begin{tabular}{lcccccccc}
    \toprule
    \textsc{\textbf{Software}} & \textsc{\textbf{P}} & \textsc{\textbf{C$\rightarrow$P}} &\textsc{\textbf{P-C}} & \textsc{\textbf{V$\rightarrow$P}} &\textsc{\textbf{P$\rightarrow$V}} &\textsc{\textbf{P-V}} &\textsc{\textbf{P-C-V}}  & \textsc{\textbf{Total}} \\
    \midrule
    HBase        & 10 & 12 & 0 & 4  & 15 & 4   &0  & 45\\  
    HDFS         & 28 & 30 & 1 & 9  & 32 & 18  &0  & 118\\ 
    Hive         & 37 & 22 & 0 & 15 & 25 & 15  &1  & 115\\
    Httpd        & 37 & 32 & 4 & 0  & 25 & 26  &1  & 125\\ 
    MapReduce    & 12 & 9  & 0 & 5  & 16 & 5   &0  & 47\\  
    MySQL        & 16 & 18 & 9 & 18 & 27 & 9   &3  & 100\\ 
    Nginx        & 37 & 12 & 10& 6  & 26 & 11  &6  & 108\\ 
    Postgres     & 15 & 17 & 3 & 2  & 16 & 8   &0  & 61 \\  
    Yarn         & 38 & 23 & 0 & 5  & 30 & 8   &0  & 104\\
    Zookeeper    & 5  & 9  & 0 & 2  & 10 & 1   &1  & 28\\   
    \midrule
    Total     & 235 & 184 & 27 & 66 & 222 & 105 & 12  & 851 \\
    \bottomrule
    \end{tabular}
    }
    \vspace{-20pt}
\end{center}
\end{table}

\subsection{Parameter interacts with Constant \& Variable (P\&C\&V)}
\label{section: PCV}

\paragraph{\textbf{Parameter, Constant and Variable take equal place (P-C-V)}}
Parameters, constants, and variables may appear in the same statement for arithmetic or logical operations. For instance, in Figure \ref{fig: nginx pcv}, the parameter \emph{keepalive\_disable}, variables \emph{safari}, and constant \emph{NGX\_HTTP\_KEEPALIVE\_DISABLE\_SAFARI} collectively determine the execution path of the branch.

\paragraph{\textbf{The combination of P\&C and P\&V}}
During the propagation, parameters may interact initially with either a constant or a variable, and subsequently with the other. 
In these scenarios, we will split the interaction into two cases, i.e., P\&C and P\&V. We will then separately investigate the interaction effects and problems of them. Here is an example in HDFS:
\begin{lstlisting}[language=myJava]
/* HDFS/RpcHandler.java */
if ((nodeLoad > maxLoad) && (maxLoad > 0)) {
  ...
  return true;
}
\end{lstlisting}

The configuration parameter \verb|maxLoad| interacts with both the constant \verb|0| and the variable \verb|nodeLoad| in the same assignment. We will split this assignment into two cases.

%% file: 4-effect.tex
\section{The effect of PCV Interaction on Software at runtime \textbf{(RQ3)}}\label{section: Interaction Effect}

\begin{figure*}
    \centerline{\includegraphics[width=\textwidth]{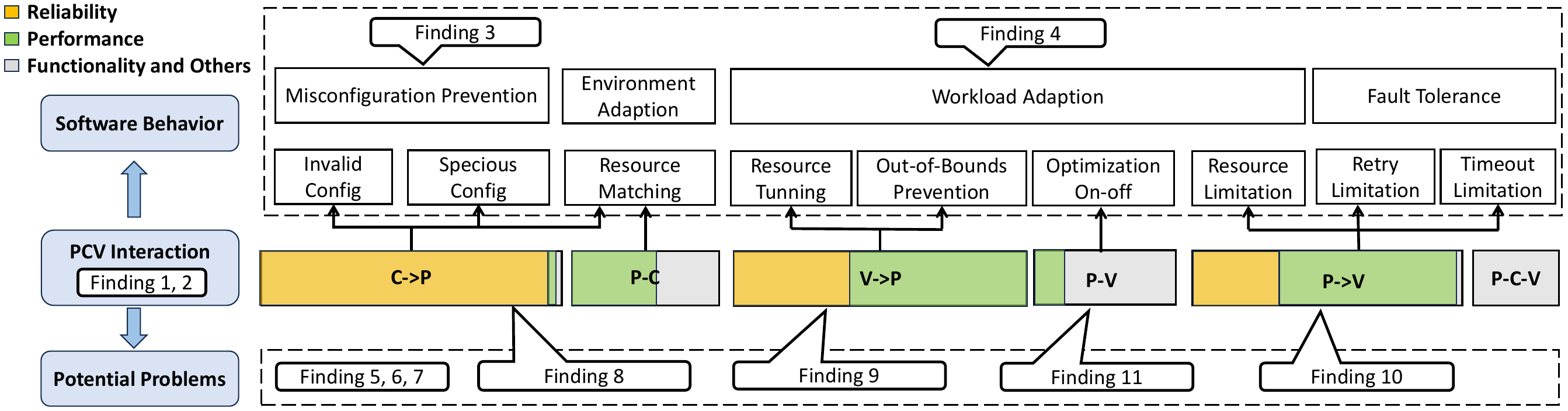}}
    \caption{PCV interaction: Patterns, Effects, and Problems (the colors represent the proportion).}
    \label{fig: sturcture}
    \vspace{-18pt}
\end{figure*}

\Space{
 \begin{center}
     \noindent\fbox{
         \parbox{.97\linewidth}
         {
             \textbf{RQ3:} How do PCV interactions take effects on software?
         }
     }
 \end{center}
 }

As PCV interactions are inevitable in software running, it is important to understand how they take effects on software at runtime. To answer this question, we first investigate the overall impact of PCV interactions, mainly from the perspective of affecting software performance and reliability (\S \ref{section: performance and reliability}). 
To achieve performance and reliability purposes, we analyze how developers manipulate PCV interactions to control software behavior at runtime (\S \ref{section: software behavoir}).

\subsection{Overall Impact} \label{section: performance and reliability}

We analyze the overall impact of PCV interaction on software from the perspective of performance and reliability, which are extremely important when providing long-term services. In this section, we study the positive sites. The potential problem will be discussed in Section~\ref{section: potential problems}. We find that 38.5\% (237/616) of the PCV interaction will affect software performance and 41.7\% (257/616) will affect reliability. While the others are usually used for functionality. 
For different interaction patterns, the impact degree is divided by three colors according to the percentage (Figure~\ref{fig: sturcture}). 


First, PCV interaction can be used to improve software performance. The following code snippet shows an example:
\begin{lstlisting}[language=myC]
/* MySQL/ddl0ctx.cc */
size_t Context::merge_io_buffer_size(...) noexcept {
  ...
  return std::max(std::max((ulong)srv_page_size,
      (ulong)IO_BLOCK_SIZE), (ulong)io_size);
}
\end{lstlisting}
The variable \emph{io\_size} records the size of the IO buffer during the file loading phase in MySQL, and \emph{IO\_BLOCK\_SIZE} (4096) is a constant value. If the user-configured value for \emph{srv\_page\_size} is too small, it may result in misalignment with the IO block and the inability to promptly process loaded files. To ensure software performance, the final size of the IO buffer will be determined by selecting the maximum value among the three.

PCV interaction can also improve software reliability. We show an example in HDFS:
\begin{lstlisting}[language=myJava]
/* HDFS/DiskBalancer.java */
if (item.getErrorCount() <= getMaxError(item))
  item.setErrMsg("Error count exceeded.");
\end{lstlisting}
The configuration parameter \emph{maxDiskErrors} controls the maximum errors allowed by the HDFS balancer when moving data blocks between two disks. 
At the branching point, the parameter interacts with \emph{getErrorCount()} to control the error count and promptly enable the replication disk, ensuring the software reliability.

\vspace{-20pt}
\begin{center}
    \noindent\begin{tcolorbox}
        \parbox{.97\linewidth}
        {
            \textbf{Finding 2: PCV interactions can have a great impact on software performance (\Space{237 / 616 = }38.5\%) and reliability (\Space{257 / 616 = }41.7\%)}. Others are usually the implementation of logical functionalities.
        }
    \end{tcolorbox}
\end{center}

\subsection{Controlling Software Behaviors} \label{section: software behavoir}

In practice, the way that PCV interaction affects performance and reliability is by controlling software behavior at runtime. 
We conclude four main aspects which are misconfiguration prevention, environment adaption, workload adaption, and fault tolerance. 
As shown in Figure \ref{fig: sturcture}, different interaction patterns can influence software behavior in multiple detailed ways. 

\subsubsection{\textbf{Misconfiguration Prevention}}
Misconfiguration consists of two parts: one involves invalid parameter values that do not meet the program's hard-coded constraints \cite{xusosp13}, and the other comprises specious parameter \cite{hu2020violet} that might lead to poor performance and reliability issues in the software. Figure~\ref{fig: httpd c->p} gives an example of using C$\rightarrow$P to prevent misconfiguration. If \emph{retry\_failover} is smaller than 0, then the method will return, preventing the invalid value from propagating and using.

We find that the most of P\&C interaction cases (81.5\%, 172/211) fall into this category. In these cases, the constants incorporate developer knowledge as they are pre-defined in the coding process. 
Developers understand the functional logic of software and the design of configuration better than users, so they are more likely to prevent misconfiguration by leveraging proper constants as checkers.

\vspace{-20pt}
\begin{center}
    \noindent\begin{tcolorbox}
        \parbox{.97\linewidth}
        {
            \textbf{Finding 3: P\&C interactions are primarily (81.5\%, 172/211) used for misconfiguration prevention.} A good constant used as a checker should incorporate developer knowledge.
        }
    \end{tcolorbox}
\end{center}



\subsubsection{\textbf{Environment Adaption}}

Parameters adapt to the environment by interacting with pre-defined constants set by developers, primarily focusing on resource matching to the hardware environment. 

Mismatching may arise between parameters and the physical hardware, such as misalignment of I/O blocks or conservative settings of I/O throughput \cite{he2023database}. To address such issues, developers proactively define constants to interact with parameters. In Figure \ref{fig: mysql p-c}, developers discovered that using half the value of parameters as the buffer size may lead to misaligned I/O blocks, resulting in performance degradation \cite{mysql-ioblock}. During maintenance, developers augmented the original buffer size with a constant, \emph{IO\_BLOCK\_SIZE}, to ensure a proper I/O block alignment.


\subsubsection{\textbf{Workload Adaption}}

The workload at runtime is constantly changing. However, parameter values will remain static if they are not updated after parsing. In this situation, developers make parameters adapt to the workload by interacting with workload-related variables. This primarily involves three modes: variables replace parameters that are unsuitable for the workload (V$\rightarrow$P), parameters and variables jointly determine the value that software uses at runtime (P-V), and parameters serve as thresholds to limit the value of variables (P$\rightarrow$V).

Take Figure \ref{fig: hive v->p.pdf} as an example, parameter \emph{threadPoolSize} controls the number of threads in partition management of Hive. If the parameter value exceeds the number of candidate tables, the size of threadpool will be determined by the value of \emph{candidates.size()}. This approach can save a certain amount of resources, enhancing the overall performance of the software.

We find a large portion of P\&V interactions (66.2\% (260/393)) fall into this category. The reason is that variables often record workload information. By interacting with these workload-related variables, configuration values can be modified to adapt to runtime workload. Specifically, we find 76.2\% (198/260) of them are to prevent excessive resource utilization. One situation is that when resources are limited, the resources beyond the limitation will restricted. Another situation is that when a task does not require too many resources, the P\&V interaction will ensure that resource allocation will not be wasted. The former ensures the reliability of the software, while the latter prevents resource-induced performance issues.

\vspace{-20pt}
\begin{center}
    \noindent\begin{tcolorbox}
        \parbox{.97\linewidth}
        {
            \textbf{Finding 4: 66.2\% (260/393) of P\&V interactions are used for workload adaption.} Among them, 79.2\% (198/260) are to prevent excessive resource utilization, which have important effects on both software performance and reliability.
        }
    \end{tcolorbox}
\end{center}

\subsubsection{\textbf{Fault Tolerance}}

When an internal error occurs, software typically waits for a certain period and then initiates a retry to maintain the availability of service. The timeout and retry number are usually determined by parameters, which are compared in conditional branches with variables representing the current retry count, constraining the software's retry attempts. Here, we show a code example:
\begin{lstlisting}[language=myJava]
/* Yarn/FederationActionRetry.java */
if (++retry > retryCount) {
  LOG.info("Maxed out Federation retries. Giving up!");
  throw e;
}
\end{lstlisting}

%% file: 5-problem.tex
\section{Potential Problems of PCV Interactions \textbf{(RQ4)}}\label{section: potential problems}

\Space{
 \begin{center}
     \noindent\fbox{
         \parbox{.97\linewidth}
         {
             \textbf{RQ4:} What are the potential problems behind PCV interactions?
         }
     }
 \end{center}
 }

A double-edged sword cuts both ways. Despite the positive effects of PCV Interactions discussed in \S~\ref{section: Interaction Effect}, we also notice that there may be some hidden issues behind those interactions. To answer RQ3, we manually analyze the source code, the commits (using \texttt{git-blame}), and issues introducing the code. We identify the risk during the process of configuration interactions (\S~\ref{section: common problems}), as well as potential problems behind specific interaction patterns (\S~\ref{section: unique problems}).

\subsection{The "Risk" under PCV Interactions}\label{section: common problems}

PCV interaction has the following three main issues: first, the configuration changes its own original effect during the interaction, which may lead to the violation of the user's intention and severe consequences (e.g., software crash). At the same time, there is a lack of logs to explicitly provide feedback to the user on key information of the interactions, which makes it difficult for users to make good adjustments to the configuration issue. Finally, we find that the software (especially Java programs) does not provide a comprehensive on-the-fly update mechanism for configuration parameters that have interactions. Thus, configuration issues in long-term services cannot be repaired at runtime, which seriously damages software reliability and availability.

\subsubsection{\textbf{Prone to bad consequences}} \label{section: bad consequence}

PCV interactions can cause bad consequences. For example in HBASE-19660 \cite{hbase19660}, when blocking store file number (V) exceeds the value of \emph{hbase.hstore.blockingStoreFiles} (P), all \texttt{write} will be stopped for more than one minute, and this can happen a few times back-to-back. 

The risks mainly exist in two ways: On the one hand, users' configured values may be overwritten (even when the value is valid), thereby violating the user's intent and producing unexpected results. 
On the other hand, if parameters fail to meet specific constraints when interacting, the interaction condition may be broken and could lead to performance degradation and runtime error. Figure~\ref{fig:intro} in the introduction gives a real-world example. Even if the parameter value is set appropriately when the service starts, the interaction relationship may be reversed as the workload increases. In all these situations, the interactions will damage the software services at runtime.  


\vspace{-25pt}
\begin{center}
    \noindent\begin{tcolorbox}
        \parbox{.97\linewidth}
        {
            \textbf{Finding 5: More than half (56.5\%, 349/616) of the interactions have the potential to cause bad consequences.} This including runtime error (\Space{48/349}13.7\%), performance degradation (\Space{186/349}53.3\%) and unexpected results (\Space{115/349}33.0\%).
        }
    \end{tcolorbox}
\end{center}

\textbf{Research Direction}: Developers and users should all be careful of PCV interactions. For developers, adequate testing for configuration interaction code should be considered. Existing fuzzing frameworks can generate configuration values as seeds to achieve higher code coverage \cite{li2024ecfuzz}. However, as PCV interactions have multiple patterns, using a simple strategy to inject configuration values cannot cover all patterns of PCV code. Future work should investigate seed mutation strategies for configuration codes and cover as many PCV codes as possible.

\subsubsection{\textbf{Lack of log information}} \label{section: lack log}
Informative log can assist users in understanding and adjusting configuration \cite{zhou2021confinlog, yuan2012conservative}. Therefore, we study whether the interaction information, especially the runtime environment/workload information will be informed to the user.

We find that, unlike error logging, logs that are related to PCV interactions are seriously inadequate. Only a small portion ($\textless$25\%) of interactions are conveyed to users through logs. Moreover, in those cases where logs are available, about a third (31.9\%) of them are checking invalid parameter values. There is rare information provided to users regarding workload-related variables. Here, we show an example of a good log:

\begin{lstlisting}[language=myJava]
/* HBase/SplitTableRegionProcedure.java */
LOG.info("pid=" + getProcId() + " splitting " + (*@\colorbox{lightred}{nbFiles}@*) + " storefiles, region=" + getParentRegion().getShortNameToLog() + ", threads=" + (*@\colorbox{lightred}{maxThreads}@*));
\end{lstlisting}
The log message explicitly points out \emph{maxThreads} (P), \emph{nbFiles} (V), and their semantic relationship. This can help users configure software better.

\vspace{-20pt}
\begin{center}
    \noindent\begin{tcolorbox}
        \parbox{.97\linewidth}
        {
            \textbf{Finding 6}: \textbf{Existing logs are lack of comprehensive interaction details (or even missing) to assist users in configuration adjustments.} Only a small number of cases (\Space{94/616}15.3\%) explicitly inform users about the interaction information.
        }
    \end{tcolorbox}
\end{center}

\textbf{Research Direction}: The same as error logging, developers should provide detailed log information about PCV interactions. 
Future work should automatically extract the interaction using program analysis. Based on the code patterns (\S~\ref{section: PCV}) and the corresponding effects (\S~\ref{section: Interaction Effect}) of the interaction code, the developers should provide detailed log information to users for configuration assistance and problem diagnosis.

\subsubsection{\textbf{Lack of on-the-fly update support}} \label{section: lack on-the-fly}
Restarting a software system that has been in service for a long time can result in significant overhead and unavoidable damage. Given the potential issues arising from PCV interactions (\S~\ref{section: bad consequence}),  we investigate whether users can mitigate and address these problems without service restart. The most effective and common approach is on-the-fly configuration update. 

We find that many parameters in C/C++ software (325/335) can be updated at runtime, corroborating the results in previous work \cite{wang2023understanding}. Surprisingly, the number is much lower in Java programs. Only 3.0\% (7/240) of parameters support on-the-fly update. In this case, it is rather difficult for users to resolve configuration issues timely and with minimal cost.


\vspace{-20pt}
\begin{center}
    \noindent\begin{tcolorbox}
        \parbox{.97\linewidth}
        {
            \textbf{Finding 7}: \textbf{The Java software does not provide sufficient support for on-the-fly update of interacted parameter.} Few of Java software parameters (7/240) support updates at runtime.
        }
    \end{tcolorbox}
\end{center}

\textbf{Research Direction}: Developers need to provide sophisticated on-the-fly update mechanism (especially for Java programs). This is to ensure configuration issues caused by interaction at runtime can be resolved without a service restart. The difficulty here may lie in how to ensure that the updated values are propagated and used correctly and are not affected by historical status.

\subsection{The "Crisis" Behind Specific Interactions}\label{section: unique problems}

The main purpose for developers to introduce PCV interactions is to help the software adapt to different environments and workloads. However, facing real complex and changing environments/workloads at runtime, developers sometimes can not guarantee the stable execution of every type of interaction, and thus introduce configuration issues. In this section we discuss the "crisis" we found behind several interaction patterns, which are: Unexplained constant used in C$\rightarrow$P, Partial adaption to workload in V$\rightarrow$P, Inappropriate threshold in P$\rightarrow$V and Missing optimization opportunities in P-V.

\subsubsection{\textbf{Unexplained constant used in C$\rightarrow$P}}

As discussed in \S~\ref{section: Interaction Effect}, C$\rightarrow$P interactions are usually used for misconfiguration prevention. In this situation, developers may conditionally overwrite the parameter value with a pre-defined constant. However, we find that those constants are not carefully selected and checked. 56.7\%, 34/60 of the constants used for parameter overwriting are not the default value (which is, theoretically the best out-of-box value for parameter). The code snippet shows an example:
\begin{lstlisting}[language=myC]
/* MySQL/srv0srv.cc */
static MYSQL_SYSVAR_ULONG(..., srv_purge_batch_size, (*@\colorbox{lightred}{300}@*), ...)  
const uint temp_batch_size = (*@\colorbox{lightred}{20}@*);

n_pages_purged = trx_purge(1,
  srv_purge_batch_size <= temp_batch_size ?
    srv_purge_batch_size : temp_batch_size,true);
\end{lstlisting}
In this case, the default value of \emph{srv\_purge\_batch\_size} is 300. However, any values greater than a "magic number" 20 will be changed to 20. Quoted from developer's comment: ``{\it a large batch size can cause a significant delay in shutdown, so reducing the batch size to magic number 20, which we hope will be sufficient to remove all the undo records .}'' \cite{mysqlmagicnumber}

We also find two cases in which parameters are mandatory overwritten by constant without conditional checking. In such cases, the parameter is a "fake" configuration. For example in \emph{srv0srv.cc} in MySQL, developers observed that user-configured values of \emph{lock\_wait\_timeout} are usually set too low. So they enforced the configuration value to one year without informing users about the change \cite{mysqllongtimeout}. This one-year timeout could lead to prolonged blocking of transactions waiting for lock resources, requiring users to employ additional deadlock detection. Ideally, the default value is the best constant to face most scenarios. 
Such inconsistency could confuse users and may violate user intention. 

\vspace{-20pt}
\begin{center}
    \noindent\begin{tcolorbox}
        \parbox{.97\linewidth}
        {
            \textbf{Finding 8:} \textbf{The constants used for parameter overwritten in C$\rightarrow$P are not carefully selected and checked.} Among the constants used for parameter overwritten, many (31/56 = 55.4\%) of them differ from the parameter's default value. There are also a few "fake configurations" that the parameter will be mandatory overwritten without checking conditions.
        }
    \end{tcolorbox}
\end{center}

\textbf{Research Direction}: Developers should be careful of using constants for parameter default value and runtime overwritten. A systematic measurement approach and test process are needed to find good constants. And making it consistent for default value and runtime parameter is overwritten. Tools should automatically find parameters that are rarely used and replace them with constants, rather than leaving a "fake" configuration.

\subsubsection{\textbf{Partial adaption to workload in V$\rightarrow$P}}

V$\rightarrow$P are usually used to adapt to runtime workload. During the interaction, developers conditionally adjust parameters according to workload-related variables. However, it falls short in two aspects. First, developers do not comprehensively consider all possible scenarios, adapting partially. Second, developers may fully respect to user-set value and only modify the values when users do not set the parameter, neglecting the necessity of runtime adaption. 

\begin{figure}
    \centerline{\includegraphics[width=0.48\textwidth]{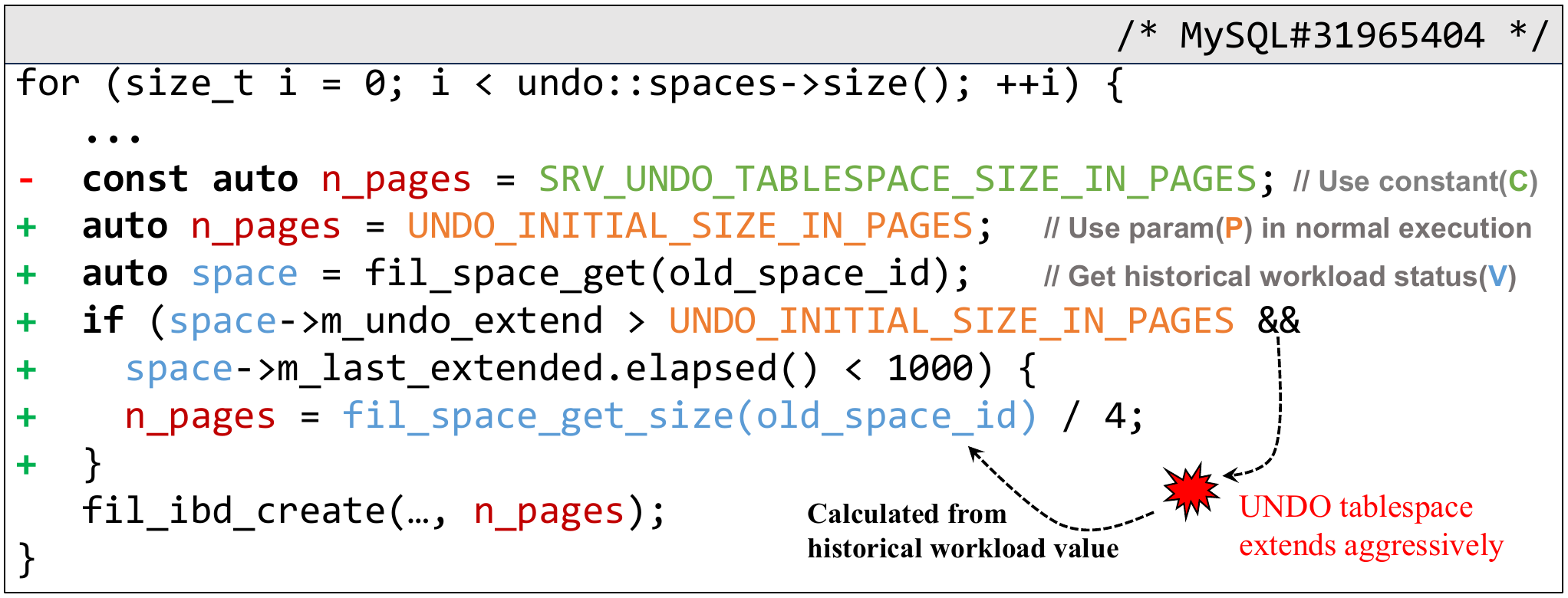}}
    \caption{Example of configuration adjustment at runtime based on historical data.}
    \label{fig: mysql dynamic}
    \vspace{-15pt}
\end{figure}

\textbf{Lack of adaption to all scenarios:}
Developers often use the \emph{max()} or \emph{min()} methods for V$\rightarrow$P interaction to select the better value. However, we find that 72.7\% (48/66) of these cases only consider one aspect (either the user-set value is too large or too small). For instance, in Figure \ref{fig: hive v->p.pdf}, the \emph{min()} method prevents the configuration value \emph{threadPoolSize} from exceeding \emph{candidates.size()} (region number) to avoid resource waste. However, if \emph{threadPoolSize} is significantly smaller than the region number, it may lead to performance issues. Conversely, while \emph{max()} can ensure sufficient resources, it may overlook cases where parameters are extensive compared to the required workload, resulting in wasted resources.

\textbf{Lack of necessary adjustment to user-set values:}
Sometimes developers are aware of the workload-related variables that will interact with parameters and have taken them into consideration. However, such modification only takes effect when users do not set parameter values at all. Here is an example:

\vspace{-5pt}
\begin{lstlisting}[language=myJava]
/* Zookeeper/NIOServerCnxnFactory.java */
int numCores = Runtime.getRuntime().availableProcessors();
numSelectorThreads = 
  Integer.getInteger(ZOOKEEPER_NIO_NUM_SELECTOR_THREADS,   
    Math.max((int) (*@\colorbox{lightred}{Math.sqrt((float) numCores / 2)}@*), 1));
\end{lstlisting}
\vspace{-5pt}

The value of the selector thread number is relevant to the number of available CPU cores. However, developers use this knowledge only when users do not set parameter \emph{ZOOKEEPER\_NIO\_NUM\_SELECTOR\_THREADS}. Even if the user sets an outrageous value, the developer will respect it here. We find 4 such cases in our dataset.

In practice, developers sometimes use historical data to adjust the configuration value, instead of blindly choosing the value set by the user \cite{mysqlnpages}. An example is shown in Figure~\ref{fig: mysql dynamic}. Developers constantly check workload-related variables (\emph{space->m\_undo\_exten} and \emph{space->m\_last\_extended.elapsed()} ) to determine whether the \texttt{UNDO tablespace} extends aggressively. If so, developers will use historical data (\emph{fil\_space\_get\_size(old\_space\_id) / 4}) as the page size rather than statically use user-set value (\emph{UNDO\_INITIAL\_SIZE\_IN\_PAGES}).

\vspace{-30pt}
\begin{center}
    \noindent\begin{tcolorbox}
        \parbox{.97\linewidth}
        {
            \textbf{Finding 9:} \textbf{During V$\rightarrow$P, developers conditionally modify user-set values to adapt to runtime workload.} However, 
            the majority (48/66=72.7\%) of conditional modifications only consider partial scenarios. Some modifications (4/66) only take effect when users do not set parameter values at all.
        }
    \end{tcolorbox}
\end{center}
\vspace{-5pt}

\textbf{Research Direction}: Developers need to make user-configured values adapt to the runtime environment and workload. A possible solution is to auto-adjust the user-set configuration value at runtime by using historical data. Such an approach requires systematic modeling and prediction of specific functions.

\subsubsection{\textbf{Inappropriate threshold in P$\rightarrow$V}}
In P$\rightarrow$V interactions, parameters typically serve as thresholds in conditional branches, which indicate whether the software is running abnormally. The inappropriate threshold can be dangerous. In HBASE-12971, when setting \emph{replication.source.maxretriesmultiplier} to 300, it will lead to a sleep time of more than 24 hours when a socket timeout exception is thrown\cite{hbase12971}.

Users often find it challenging to set such parameter values as it is difficult to predict runtime workload. For instance, the parameter \emph{thread\_stack} in MySQL on a 32-bit platform has a valid range of 128 KB to 4 GB. However, users set it to 192 KB, and the MySQL server encounters a "thread stack overrun" \cite{mysqlstack}. The user changes it to 256KB and solves the problem. 

\begin{figure}
    \centerline{\includegraphics[width=0.48\textwidth]{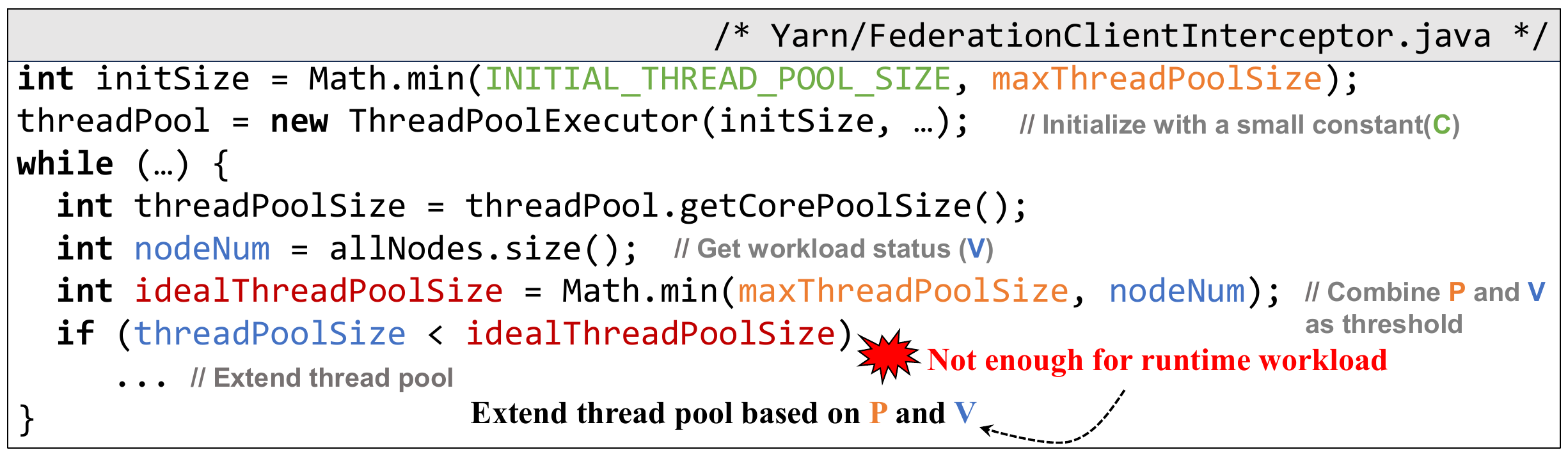}}
    \caption{Example of elastic threshold.}
    \label{fig: yarn dynamic}
    \vspace{-15pt}
\end{figure}

Figure~\ref{fig: yarn dynamic} shows a good practice in Yarn that makes the threshold elastic at runtime. In this case, the threshold of the thread pool (\emph{idealThreadPoolSize}) is not only determined by users. Rather, it uses a user-set threshold (\emph{maxThreadPoolSize}) and workload-related variable (\emph{nodeNum}) to co-determine the threshold. When the user sets \emph{maxThreadPoolSize} too large, it will be restricted by (\emph{nodeNum}). This is actually a case that uses a V$\rightarrow$P interaction to constrain P and then uses P as a threshold in another P$\rightarrow$V interaction to control software behavior. Such a combination takes the advantage of balancing user intention with actual software requirements.

\vspace{-20pt}
\begin{center}
    \noindent\begin{tcolorbox}
        \parbox{.97\linewidth}
        {
            \textbf{Finding 10:} \textbf{For parameters serving as thresholds in P$\rightarrow$V, they themselves also need appropriate limitations.} Arbitrarily setting their values without thorough consideration of the runtime scenario can lead to severe issues.
        }
    \end{tcolorbox}
\end{center}

\textbf{Research Direction}: Automated tools are needed to make good predictions of threshold or make threshold elastic at runtime by using workload information to constrain threshold-type parameters. Such an operation needs the combination of multiple P\&V interactions.

\subsubsection{\textbf{Missing optimization opportunities in P-V}}
Besides functionality, P-V interactions are used for optimization tuning. Unlike a simple switch parameter, the workload-related variables in P-V interaction usually represent the requirement of optimization. For example in Figure \ref{fig: hdfs p-v}, when \emph{stageSubSections.size()} is positive, which means there are multiple subsections in the corresponding file system image to process, the configuration value \emph{loadInParallel} should be on to execute parallel loading. However, the default value is false, increasing the possibility of missing optimization even when needed.

When developers introduce a new or advanced feature, they often turn it off for the conservative principles \cite{zhang2021evolutionary, he2022multi, he2020cp-detector}. After using it for a long time in production, they may change the default value to true. For example, in HDFS-12603~\cite{HDFS-12603}, Developers turn on async edit logging as default after ``{\it running it in production for quite a while with no issues.}''

\vspace{-20pt}
\begin{center}
    \noindent\begin{tcolorbox}
        \parbox{.97\linewidth}
        {
            \textbf{Finding 11:} P-V demonstrates the need for tuning switches. A simple switch-off could lose the opportunity for performance optimization even when needed. \Space{Nearly all}Many of the default values (14/21=66.7\%) of parameters used for tuning optimization are false. 
        }
    \end{tcolorbox}
\end{center}

\textbf{Research Direction}: Future work should monitor the optimization-related variable and design strategies to automatically suggest the scenarios in which the user should turn on the optimization parameter.

%% file: 7-threats.tex
\section{Threats to validity} \label{section: threat}
We mainly choose mature data-intensive software systems, which are widely deployed, highly configurable, and well-documented. What's more, they have a rich development and maintenance history (on GitHub and JIRA). Although we choose different types of software, our findings may not apply to all kinds of software. Besides, even random sampling might introduce bias, potentially affecting the results. However, we still believe that our study is representative enough to reveal the advantages and potential problems of the PCV interactions as we did not restrict our research to a certain type of software or a specific value type of configuration parameters. Static analysis tools may miss some propagation paths due to program complexity. To ensure completeness, we manually examine the code to identify PCV interactions. While we can't guarantee capturing all cases, our dataset includes all interaction types with enough instances for a thorough study.

%% file: 8-related.tex
\section{Related work}
\textbf{Configuration design and comprehension.} Some works focus on understanding and improving software configuration design\cite{xu2015hey, zhang2021evolutionary, sayagh2018software}. The work of Zhang et al~\cite{zhang2021evolutionary} closely aligns with our study. They study the design and implementation of software configurations at the source code level. Our research is similarly rooted in the source code but diverges by emphasizing the influence of configurations on software behavior. Sayagh et al. \cite{sayagh2018software} summarized common activities in software configuration engineering undertaken by developers through literature reviews and expert interviews. Some research has focused on the design and usage of Software Product Line (SPL) configurations~\cite{soares2018exploring, nadi2013study}, predominantly employed by developers to facilitate collaborative development and testing. Our work focuses on how configuration parameters affect software behavior at the source code level, but not the activities of the configuration management.


\textbf{Misconfiguration detection and prevention.} There has been a substantial body of studies on misconfiguration \cite{2021fb10, 2021fbWhatsApp, gunawi2016does, maurer2015fail, nagaraja2004understanding, oppenheimer2003internet, rabkin2011precomputing, rabkin2012hadoop, tang2015holistic, yin2011empirical}. Many works focus on preventing misconfiguration before software execution \cite{xu2016early, xusosp13, cdep, sun2020testing}. There are also works dedicated to detecting configuration issues at runtime \cite{huang2015confvalley, yuan2011context, wang2023understanding, zhou2023wmwatcher}. Our work neither focuses on detecting misconfigurations nor on diagnosing software issues caused by misconfigurations. Instead, we concentrate on the interaction of configuration parameters with constants and variables in the source code. We aim to gain a comprehensive and deep understanding of how runtime configuration affects software behavior and use the knowledge to help developers and users design/use configuration better. 

%% file: 9-conclusion.tex
\section{Conclusion}


    This paper presents, to the best of our knowledge, the first comprehensive study on how configurations affect software at runtime by focusing on the interaction between configuration parameters, constants, and variables (PCV Interaction). This interaction reflects how user intentions, developer knowledge, and runtime environment/workload jointly influence software behavior.
    We analyze 705 configuration parameters from 10 large-scale software systems, revealing that 66.4\% of these parameters interact with constants or variables after parsing. We categorize these interactions and examine their runtime effects, highlighting the risks and potential problems associated with specific interaction patterns. Our findings offer new insights and encourage the development of automated techniques to improve both software configuration by users and configuration design by developers.

%% file: 10-acknowledege.tex
\section*{ACKNOWLEDGMENTS}
We would like to thank the anonymous reviewers for their
insightful comments. This research was funded by NSFC No. 62272473, the Science and Technology Innovation Program of Hunan Province (No.2023RC1001), and NSFC No.62202474.

%% file: 0-main.bbl
\begin{thebibliography}{10}
\providecommand{\url}[1]{#1}
\csname url@samestyle\endcsname
\providecommand{\newblock}{\relax}
\providecommand{\bibinfo}[2]{#2}
\providecommand{\BIBentrySTDinterwordspacing}{\spaceskip=0pt\relax}
\providecommand{\BIBentryALTinterwordstretchfactor}{4}
\providecommand{\BIBentryALTinterwordspacing}{\spaceskip=\fontdimen2\font plus
\BIBentryALTinterwordstretchfactor\fontdimen3\font minus \fontdimen4\font\relax}
\providecommand{\BIBforeignlanguage}[2]{{%
\expandafter\ifx\csname l@#1\endcsname\relax
\typeout{** WARNING: IEEEtran.bst: No hyphenation pattern has been}%
\typeout{** loaded for the language `#1'. Using the pattern for}%
\typeout{** the default language instead.}%
\else
\language=\csname l@#1\endcsname
\fi
#2}}
\providecommand{\BIBdecl}{\relax}
\BIBdecl

\bibitem{zhang2021evolutionary}
Y.~Zhang, H.~He, O.~Legunsen, S.~Li, W.~Dong, and T.~Xu, ``An evolutionary study of configuration design and implementation in cloud systems,'' in \emph{2021 IEEE/ACM 43rd International Conference on Software Engineering (ICSE)}.\hskip 1em plus 0.5em minus 0.4em\relax IEEE, 2021, pp. 188--200.

\bibitem{wang2023understanding}
T.~Wang, Z.~Jia, S.~Li, S.~Zheng, Y.~Yu, E.~Xu, S.~Peng, and X.~Liao, ``Understanding and detecting on-the-fly configuration bugs,'' in \emph{Proceedings of the 45th International Conference on Software Engineering (ICSE)}, 2023.

\bibitem{yin2011empirical}
Z.~Yin, X.~Ma, J.~Zheng, Y.~Zhou, L.~N. Bairavasundaram, and S.~Pasupathy, ``An empirical study on configuration errors in commercial and open source systems,'' in \emph{Proceedings of the 23th ACM Symposium on Operating Systems Principles (SOSP'11)}, 2011, pp. 159--172.

\bibitem{xusosp13}
T.~Xu, J.~Zhang, P.~Huang, J.~Zheng, T.~Sheng, D.~Yuan, Y.~Zhou, and S.~Pasupathy, ``Do not blame users for misconfigurations,'' in \emph{Proceedings of the 24th ACM Symposium on Operating Systems Principles (SOSP'13)}, 2013, pp. 244,259.

\bibitem{xu2015hey}
T.~Xu, L.~Jin, X.~Fan, Y.~Zhou, S.~Pasupathy, and R.~Talwadker, ``Hey, you have given me too many knobs!: Understanding and dealing with over-designed configuration in system software,'' in \emph{Proceedings of the 2015 10th Joint Meeting on Foundations of Software Engineering}, 2015, pp. 307--319.

\bibitem{2021fbWhatsApp}
P.~Wallpapers, ``Facebook, instagram, whatsapp hit by global outage,'' https://www.pukmedia.com/EN/Details/69501, 2021.

\bibitem{xu2016early}
T.~Xu, X.~Jin, P.~Huang, Y.~Zhou, S.~Lu, L.~Jin, and S.~Pasupathy, ``Early detection of configuration errors to reduce failure damage,'' in \emph{12th USENIX Symposium on Operating Systems Design and Implementation (OSDI'16)}, 2016, pp. 619--634.

\bibitem{li2021challenges}
W.~Li, Z.~Jia, S.~Li, Y.~Zhang, T.~Wang, E.~Xu, J.~Wang, and X.~Liao, ``Challenges and opportunities: an in-depth empirical study on configuration error injection testing,'' in \emph{Proceedings of the 30th ACM SIGSOFT International Symposium on Software Testing and Analysis}, 2021, pp. 478--490.

\bibitem{HBASE-21000}
``Hbase-21000. default limits for pressureawarecompactionthroughputcontroller are too low.'' https://issues.apache.org/jira/browse/HBASE-21000, 2018.

\bibitem{HBASE24544}
``Hbase-24544. recommend upping zk jute.maxbuffer in all but minor installs.'' https://issues.apache.org/jira/browse/HBASE-24544, 2020.

\bibitem{hbase12971}
``Hbase-12971. replication stuck due to large default value for replication.source.maxretriesmultiplier.'' https://issues.apache.org/jira/browse/HBASE-12971, 2015.

\bibitem{hbase19660}
``Hbase-19660. up default retries from 10 to 15 and blocking store files limit from 10 to 16,'' https://issues.apache.org/jira/browse/HBASE-19660, 2017.

\bibitem{HDFS-12603}
``Hdfs-12603. enable async edit logging by default,'' https://issues.apache.org/jira/browse/HDFS-12603, 2017.

\bibitem{mysqlstack}
S.~Overflow, ``Mysql server's thread\_stack parameter - what is it? how big should it be?'' https://stackoverflow.com/questions/2919558/mysql-servers-thread-stack-parameter-what-is-it-how-big-should-it-be, 2010.

\bibitem{huang2015confvalley}
P.~Huang, W.~J. Bolosky, A.~Singh, and Y.~Zhou, ``Confvalley: A systematic configuration validation framework for cloud services,'' in \emph{Proceedings of the Tenth European Conference on Computer Systems}, 2015, pp. 1--16.

\bibitem{zhang2014encore}
J.~Zhang, L.~Renganarayana, X.~Zhang, N.~Ge, V.~Bala, T.~Xu, and Y.~Zhou, ``Encore: Exploiting system environment and correlation information for misconfiguration detection,'' in \emph{Proceedings of the 19th international conference on Architectural support for programming languages and operating systems}, 2014, pp. 687--700.

\bibitem{he2020cp-detector}
H.~He, Z.~Jia, S.~Li, E.~Xu, T.~Yu, Y.~Yu, J.~Wang, and X.~Liao, ``Cp-detector: Using configuration-related performance properties to expose performance bugs,'' in \emph{Proceedings of the 35th IEEE/ACM International Conference on Automated Software Engineering}, 2020, pp. 623--634.

\bibitem{sun2020testing}
X.~Sun, R.~Cheng, J.~Chen, E.~Ang, O.~Legunsen, and T.~Xu, ``Testing configuration changes in context to prevent production failures,'' in \emph{14th USENIX Symposium on Operating Systems Design and Implementation (OSDI 20)}, 2020, pp. 735--751.

\bibitem{conftainter}
T.~Wang, H.~He, X.~Liu, S.~Li, Z.~Jia, Y.~Jiang, Q.~Liao, and W.~Li, ``Conftainter: Static taint analysis for configuration options,'' in \emph{2023 38th IEEE/ACM International Conference on Automated Software Engineering (ASE 2023)}, 2023, pp. 1640--1651.

\bibitem{cflow}
``Cflow,'' https://github.com/xlab-uiuc/cflow, 2024.

\bibitem{jira}
``Jira,'' https://issues.apache.org/jira/issues/, 2024.

\bibitem{github}
``Github,'' https://github.com/, 2024.

\bibitem{cdep}
Q.~Chen, T.~Wang, O.~Legunsen, S.~Li, and T.~Xu, ``{Understanding and Discovering Software Configuration Dependencies in Cloud and Datacenter Systems},'' in \emph{Proceedings of the 2020 ACM Joint European Software Engineering Conference and Symposium on the Foundations of Software Engineering (ESEC/FSE'20)}, November 2020.

\bibitem{hu2020violet}
Y.~Hu, G.~Huang, and P.~Huang, ``Automated reasoning and detection of specious configuration in large systems with symbolic execution,'' in \emph{14th USENIX Symposium on Operating Systems Design and Implementation (OSDI 20)}, 2020, pp. 719--734.

\bibitem{he2023database}
H.~He, E.~Xu, S.~Li, Z.~Jia, S.~Zheng, Y.~Yu, J.~Ma, and X.~Liao, ``When database meets new storage devices: Understanding and exposing performance mismatches via configurations,'' \emph{Proceedings of the VLDB Endowment}, vol.~16, no.~7, pp. 1712--1725, 2023.

\bibitem{mysql-ioblock}
``Bug \#33570629 [innodb] mysql server crash,'' https://github.com/mysql/mysql-server/commit/1b900503e03d, 2021.

\bibitem{li2024ecfuzz}
J.~Li, S.~Li, K.~Li, F.~Luo, H.~Yu, S.~Li, and X.~Li, ``Ecfuzz: Effective configuration fuzzing for large-scale systems,'' in \emph{Proceedings of the 46th IEEE/ACM International Conference on Software Engineering}, 2024, pp. 1--12.

\bibitem{zhou2021confinlog}
S.~Zhou, X.~Liu, S.~Li, Z.~Jia, Y.~Zhang, T.~Wang, W.~Li, and X.~Liao, ``Confinlog: Leveraging software logs to infer configuration constraints,'' in \emph{2021 IEEE/ACM 29th International Conference on Program Comprehension (ICPC)}.\hskip 1em plus 0.5em minus 0.4em\relax IEEE, 2021, pp. 94--105.

\bibitem{yuan2012conservative}
D.~Yuan, S.~Park, P.~Huang, Y.~Liu, M.~M. Lee, X.~Tang, Y.~Zhou, and S.~Savage, ``Be conservative: Enhancing failure diagnosis with proactive logging,'' in \emph{10th USENIX Symposium on Operating Systems Design and Implementation (OSDI 12)}, 2012, pp. 293--306.

\bibitem{mysqlmagicnumber}
``Bug \#21040050 purge thread must exit sooner at server shutdown,'' https://github.com/mysql/mysql-server/commit/3e8202ff4439, 2015.

\bibitem{mysqllongtimeout}
``Followup to bug\#45225 locking: hang if drop table with no timeout,'' https://github.com/mysql/mysql-server/commit/dd42aab8406e, 2010.

\bibitem{mysqlnpages}
``Bug\#31965404: Tps regression when automatic undo truncate is enabled,'' https://github.com/mysql/mysql-server/commit/60fd370cae5a, 2020.

\bibitem{he2022multi}
H.~He, Z.~Jia, S.~Li, Y.~Yu, C.~Zhou, Q.~Liao, J.~Wang, and X.~Liao, ``Multi-intention-aware configuration selection for performance tuning,'' in \emph{Proceedings of the 44th International Conference on Software Engineering}, 2022, pp. 1431--1442.

\bibitem{sayagh2018software}
M.~Sayagh, N.~Kerzazi, B.~Adams, and F.~Petrillo, ``Software configuration engineering in practice interviews, survey, and systematic literature review,'' \emph{IEEE Transactions on Software Engineering}, vol.~46, no.~6, pp. 646--673, 2018.

\bibitem{soares2018exploring}
L.~R. Soares, J.~Meinicke, S.~Nadi, C.~K{\"a}stner, and E.~S. de~Almeida, ``Exploring feature interactions without specifications: A controlled experiment,'' \emph{ACM SIGPLAN Notices}, vol.~53, no.~9, pp. 40--52, 2018.

\bibitem{nadi2013study}
S.~Nadi, ``A study of variability spaces in open source software,'' in \emph{2013 35th International Conference on Software Engineering (ICSE)}.\hskip 1em plus 0.5em minus 0.4em\relax IEEE, 2013, pp. 1353--1356.

\bibitem{2021fb10}
S.~Janardhan, ``Update about the october 4th outage,'' https://engineering.fb.com/2021/10/04/networking-traffic/outage/, 2021.

\bibitem{gunawi2016does}
H.~S. Gunawi, M.~Hao, R.~O. Suminto, A.~Laksono, A.~D. Satria, J.~Adityatama, and K.~J. Eliazar, ``Why does the cloud stop computing? lessons from hundreds of service outages,'' in \emph{Proceedings of the Seventh ACM Symposium on Cloud Computing}, 2016, pp. 1--16.

\bibitem{maurer2015fail}
B.~Maurer, ``Fail at scale: Reliability in the face of rapid change,'' \emph{Queue}, vol.~13, no.~8, pp. 30--46, 2015.

\bibitem{nagaraja2004understanding}
K.~Nagaraja, F.~Oliveira, R.~Bianchini, R.~P. Martin, and T.~D. Nguyen, ``Understanding and dealing with operator mistakes in internet services.'' in \emph{OSDI}, vol.~4, 2004, pp. 61--76.

\bibitem{oppenheimer2003internet}
D.~Oppenheimer, A.~Ganapathi, and D.~A. Patterson, ``Why do internet services fail, and what can be done about it?'' in \emph{4th Usenix Symposium on Internet Technologies and Systems (USITS 03)}, 2003.

\bibitem{rabkin2011precomputing}
A.~Rabkin and R.~Katz, ``Precomputing possible configuration error diagnoses,'' in \emph{2011 26th IEEE/ACM International Conference on Automated Software Engineering (ASE 2011)}.\hskip 1em plus 0.5em minus 0.4em\relax IEEE, 2011, pp. 193--202.

\bibitem{rabkin2012hadoop}
A.~Rabkin and R.~H. Katz, ``How hadoop clusters break,'' \emph{IEEE software}, vol.~30, no.~4, pp. 88--94, 2012.

\bibitem{tang2015holistic}
C.~Tang, T.~Kooburat, P.~Venkatachalam, A.~Chander, Z.~Wen, A.~Narayanan, P.~Dowell, and R.~Karl, ``Holistic configuration management at facebook,'' in \emph{Proceedings of the 25th symposium on operating systems principles}, 2015, pp. 328--343.

\bibitem{yuan2011context}
D.~Yuan, Y.~Xie, R.~Panigrahy, J.~Yang, C.~Verbowski, and A.~Kumar, ``Context-based online configuration-error detection,'' in \emph{Proceedings of the 2011 USENIX conference on USENIX annual technical conference}, 2011, pp. 28--28.

\bibitem{zhou2023wmwatcher}
S.~Zhou, Z.~Jiang, S.~Li, X.~Liu, Z.~Jia, Y.~Zhang, J.~Ma, and H.~Mi, ``Wmwatcher: Preventing workload-related misconfigurations in production environment,'' in \emph{2023 30th Asia-Pacific Software Engineering Conference (APSEC)}.\hskip 1em plus 0.5em minus 0.4em\relax IEEE, 2023, pp. 279--288.

\end{thebibliography}
